\documentclass{iauc}
\usepackage{graphicx}

\title[Dwarf spheroidals in nearby clusters] 
{The counterparts of Local Group dwarf spheroidals in nearby clusters}

\author[M. Hilker, S. Mieske \& L. Infante]   
{Michael Hilker$^1$, Steffen Mieske$^1$ \break \and L. Infante$^2$}

\affiliation{$^1$Sternwarte der Universit\"at Bonn, Auf dem H\"ugel 71,
53121 Bonn, Germany \break 
email: mhilker@astro.uni-bonn.de, smieske@astro.uni-bonn.de\\[\affilskip]
$^2$Departamento de Astronom\'\i a y Astrof\'\i sica, P.~Universidad 
Cat\'olica, \break Casilla 104, Santiago 22, Chile \break
email: linfante@astro.puc.cl}

\pubyear{2005}
\volume{198} 
\pagerange{??--??}
\date{?? and in revised form ??}
\jname{Near-field cosmology with dwarf elliptical galaxies}
\editors{H. Jerjen and B. Binggeli, eds.}
\begin{document}

\maketitle

\begin{abstract}
In this contribution, we report on the discovery of dwarf spheroidals (dSphs) 
in the core of the Fornax cluster. Their photometric properties -- like 
magnitude, colour, surface brightness -- are very similar to 
those of Local Group dSphs. However, at a given total magnitude, 
dSphs in Fornax seem to be more extended than their Local Group counterparts.
The membership of several dwarf galaxy candidates in Fornax has been confirmed
by surface brightness fluctuation measurements on deep wide-field images taken 
with the Magellan telescope. The analysis of these images also confirms
the flat faint end slope of the luminosity function for dSphs
in Fornax which contradicts the expected large number 
of small dark matter halos connected to dwarf galaxies in $\Lambda$CDM theory.
Dwarf spheroidals have also been detected in the Hydra\,I and Centaurus 
cluster. A preliminary analysis of their photometric properties shows that
they obey similar scaling relations
as their counterparts in Fornax and the Local Group.

\end{abstract}

\firstsection 
             
\section{Introduction}

The identification of dwarf galaxies in different environments plays an 
important role for the verification of cosmological models. Cold dark matter
models predict a large number of small dark matter (DM) halos as sub-structures 
around Milky Way-sized halos as well as galaxy cluster-sized halos (e.g. Moore 
et al. \cite{moor99}). If every small DM halo would contain luminous matter, 
many dwarf galaxies are expected in basically all environments. However,
already in the Local Group there exists a strong discrepancy, the so-called
``missing satellite problem'' (e.g. Klypin et al. \cite{klyp99}). The faint end
of its galaxy luminosity function is quite flat ($\alpha=-1.1$, Pritchet \& 
van den Bergh \cite{prit99}) compared with CDM model predictions. 
Several studies of galaxy clusters show rising numbers of low-luminosity
dEs down to about $M_V\simeq-11$ mag with very different faint-end slopes in
the range $-1.1>\alpha>-2.2$ (e.g. Trentham \& Tully \cite{tren02}). However, 
it is not known whether the luminosity function in clusters continues 
to $M_V=-8.5$, as it does in the Local Group. Do dSphs actually exist in galaxy 
clusters? And how abundant are they?

Here, we present the identification and analysis of dwarf spheroidals in
nearby clusters, namely the Fornax, Hydra\,I and Centaurus cluster.

\begin{figure}
\centering
\resizebox{6.7cm}{!}{\includegraphics{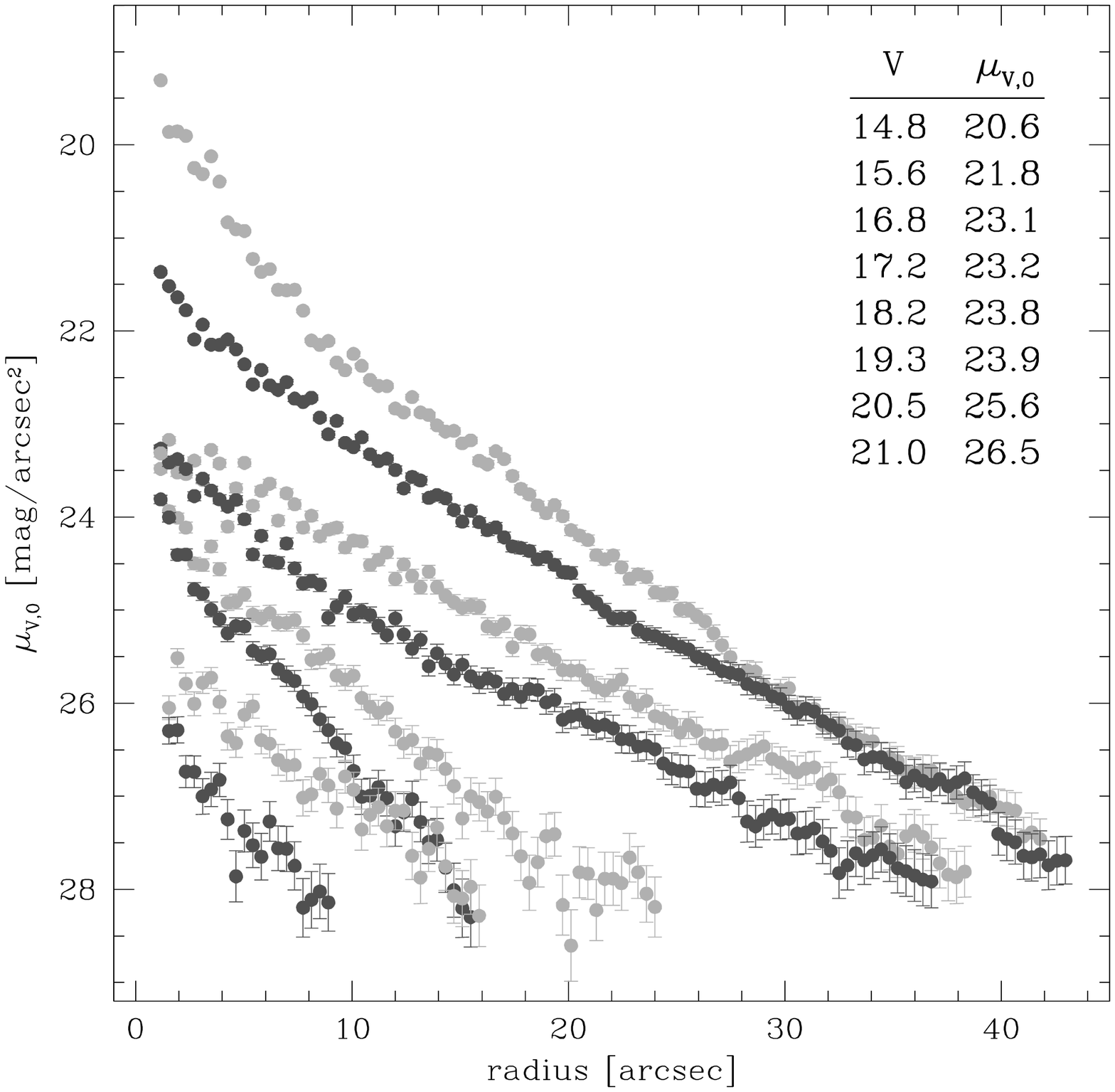} }
\resizebox{6.7cm}{!}{\includegraphics{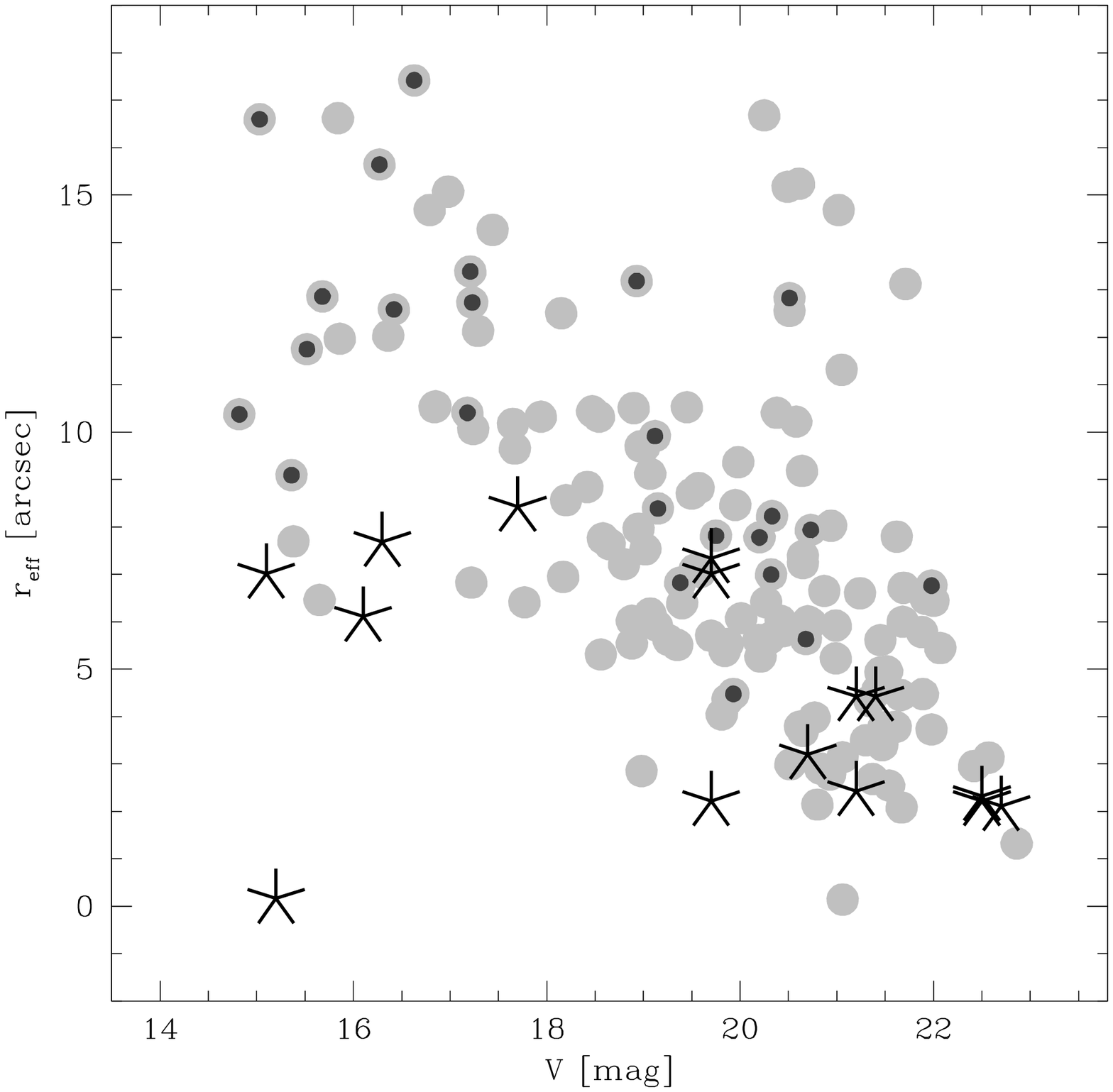} }
\caption{{\bf Left:} Surface brightness profiles of several dEs and dSphs in
the Fornax cluster. The total apparent $V$ magnitudes and central surface
brightnesses (exponential law) are indicated. {\bf Right:} Scale lengths of
exponential fits to the outer parts of the surface brightness profiles for
dSphs/dEs in Fornax (grey circles) and the Local Group (asterisks) versus $V$
magnitude.}
\end{figure}

\section{WFCCD survey of dwarf spheroidals in the Fornax cluster core}

End of 1999, we observed the core of the Fornax cluster through the Johnson 
$VI$ filters with the wide-field CCD (WFCCD) at the 100-inch du~Pont telescope 
at Las Campanas, Chile.  The pixel scale of 0.77 arcsec/pixel favours the 
detection of very low surface brightness galaxies.
The photometric parameters of the dEs/dSphs have been derived from the
analysis of their surface brightness profiles (see Fig.~1): the total magnitude
by a curve of growth analysis, the color within an aperture of 8 arcsec 
diameter, and the central surface brightness from an exponential fit to the 
outer part of the profile.

First results of our study were presented in Hilker et al. 
(\cite{hilk03}).
About 70 new dSph candidates have been discovered beyond the limits
of the Fornax Cluster Catalog (FCC, Ferguson \cite{ferg89}), the faintest one
with an absolute magnitude of $M_V\simeq-8.8$ mag and a central surface
brightness of $\mu_V\simeq27$ mag/arcsec$^2$. The dSphs follow similar
magnitude-surface brightness and also colour-magnitude relations as their 
counterparts in the Local Group (see Figs.~1 and 2 in Hilker et al. 
\cite{hilk03}).
The faint-end slope of the luminosity function of the dSphs is 
flat ($\alpha\simeq-1.1\pm0.1$, see Fig.~3 in Hilker et al. \cite{hilk03}).

The only difference between the Local Group dSphs and those in the Fornax 
cluster are their sizes. Fig.~2 shows that at a given total magnitude, 
the dEs/dSphs in Fornax are on average larger than their Local Group 
counterparts (data from Grebel et al. \cite{greb03}).
This is also reflected in the magnitude-surface brightness diagram (Fig.~3)
where the Local Group dwarfs have on average a higher central surface
brightness than the Fornax dSphs. This finding might be explained by 
environmental effects, i.e. the strong tidal forces in a cluster that lead to 
more extended, lower surface brightness dwarf galaxies, and finally to a 
destruction of dSphs close to the cluster center (e.g.  Hilker et al. 
\cite{hilk99}).

\begin{figure}
\centering
\includegraphics[width=1.0\textwidth]{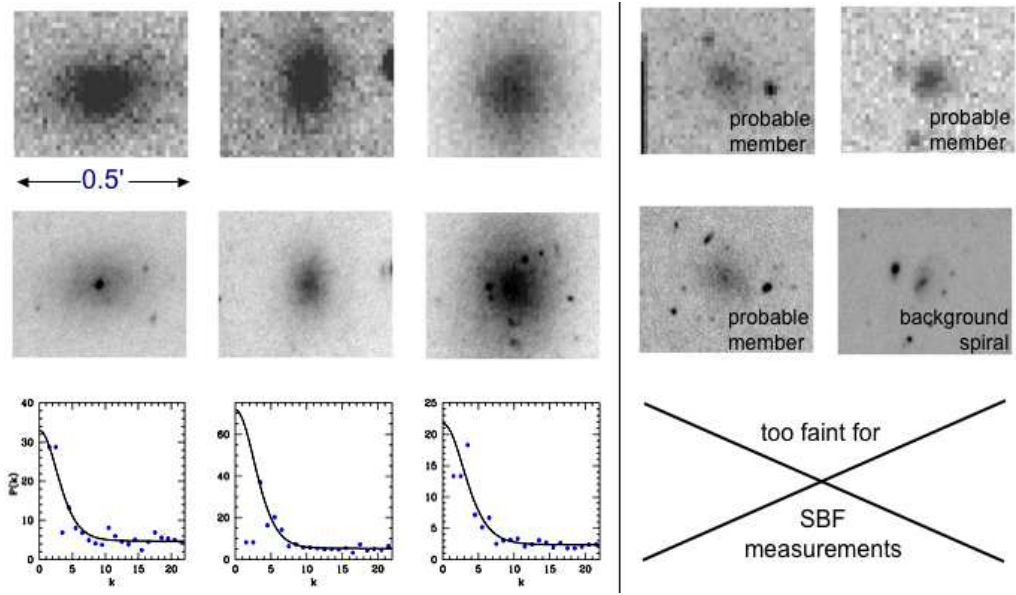}
\caption{{\bf Left:} three examples for SBF measurements in dEs. The top and 
middle row show thumbnail images galaxies from the WFCCD and IMACS data sets, 
respectively. In the bottom row the azimuthally averaged power spectra of the 
SBF measurement and the corresponding fits are plotted. 
{\bf Right:} four galaxies are shown that are too faint or small for SBF 
measurements.}
\end{figure}

\section{SBF measurements in the Fornax Deep Field}

The WFCCD study has shown that the combination of deep multi-color photometry
in a wide field with a sufficient resolution is crucial in order to
identify dwarf spheroidal candidates in nearby clusters.
However, one still can not be sure that all dSph candiates are real cluster 
members, as long as one has not measured their distances.

Therefore, to improve the membership assignment, we have re-imaged the central 
Fornax cluster with substantially better spatial resolution (0.2 arcsec/pixel)
using IMACS and Magellan at LCO. From the first part of our 
survey, we directly determined the cluster 
membership for 10 previously unconfirmed candidate dEs in the magnitude 
range $-14.2<M_V<- 11$ mag using the SBF method (Fig.~2, see Mieske et al. 
\cite{mies03a} for details of this method).
This extends the magnitude range of confirmed cluster members far into the 
regime where the faint end slope $\alpha$ dominates the galaxy luminosity 
function. 

Furthermore, we improved the morphological cluster membership assignments for 
fainter galaxies ($M_V<-10$ mag). For the vast majority of dE candidates we 
confirm the probable cluster membership, such that $\alpha$ changes by less 
than 0.02 compared to the previously determined value. Only very few dSph 
candidates turn out te be likely background galaxies due to their better
resolved morphology (see Fig.~2). We find two new dSph 
candidates from our IMACS imaging (see Fig.~3). Including them does not change 
$\alpha$, either. This further confirms the strong 
discrepancy between the number of low mass dark matter halos expected in a 
$\Lambda$CDM universe and the number of low luminosity galaxies. 

The new membership determinations from the first part of the ``Fornax Deep 
Field'' survey are summarised in the magnitude-surface brightness plot 
(see Fig.~3). Except for a few barely resolved galaxies (close to the dashed 
line), the morphological membership classification seems to work very reliable.

\begin{figure}
\centering
\includegraphics[width=0.78\textwidth]{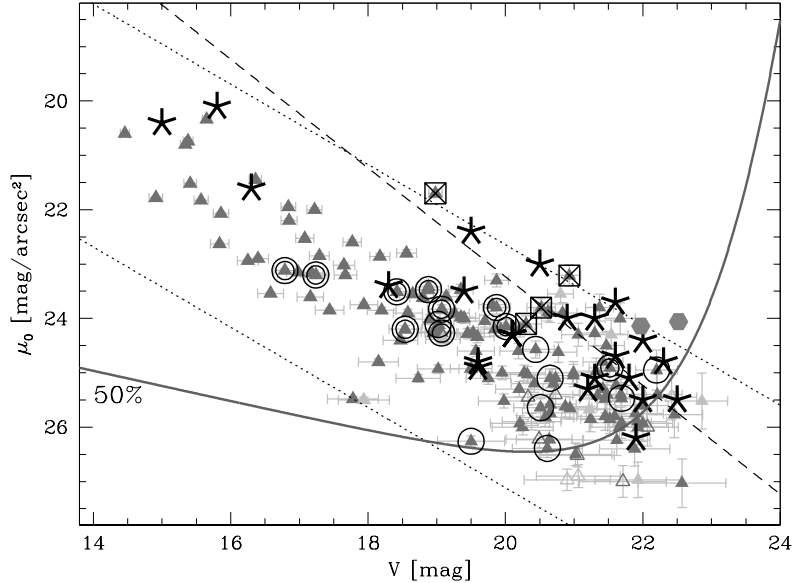}
\caption{Magnitude-surface brightness diagram of dEs and dSphs in the Fornax
cluster (triangles) and the Local Group (asterisks). Light grey triangles mark
galaxies that lie outside 3$\sigma$ of the colour-magnitude relation, open
triangles those galaxies that were identified only by eye. Double circles
indicate cluster membership confirmed with SBF, single circles probable
membership due to morphology. Probable background galaxies (due to morphology)
are crossed out. The two filled hexagons mark two additional cluster member
candidates found from the IMACS imaging. The dashed line
indicates a scale length of 2.5 arcsec for an exponential profile. The dotted
lines are the 3$\sigma$ deviations from the magnitude-surface brightness
relation as defined by the Fornax dEs. The solid curve shows the completeness
limit of 50\%.}
\end{figure}

\section{Dwarf spheroidals in the Hydra\,I and Centaurus cluster}

The central parts of the Hydra\,I and Centaurus cluster have been imaged
through Johnson $VI$ with FORS1 at the VLT in dark time and under excellent
seeing conditions (for first results, see Mieske \& Hilker \cite{mies03b}). A
first visual inspection of the images revealed a wealth of faint, spatially
resolved dSph candidates. The surface brightness profiles of a small
random sample has been measured. Their structural and photometric parameters 
are such that they fall on top of the colour-magnitude and magnitude-surface 
brightness relation defined by the Fornax dSphs. The publication of these
results is in preparation.

\section{Conclusions}

\noindent
{\bf 1.} The counterparts of Local Group dSphs do exist in Fornax. They follow 
similar colour-magnitude and magnitude-surface brightness relations, except 
that the Fornax dSphs seem to be in average larger at a given total
magnitude than the Local Group dSphs.\\
{\bf 2.} SBF-cluster memberships and improved morphological classifications do 
not change the shallow faint end slope of $\alpha \simeq -1.1$ derived from the 
WFCCD-data by more than 0.02, posing a challenge to $\Lambda$CDM theory.\\
{\bf 3.} Preliminary analysis of deep VLT images in Hydra\,I and Centaurus show
that dSph candidates exist in these clusters, and that they follow similar
scaling relations as their counterparts in Fornax. More on this in forthcoming
publications.

%
%
%

\end{document}